\documentclass{article}
\usepackage{PRIMEarxiv}

\usepackage[utf8]{inputenc} 
\usepackage[T1]{fontenc}    
\usepackage{hyperref}       
\usepackage{url}            
\usepackage{booktabs}       
\usepackage{amsfonts}       
\usepackage{nicefrac}       
\usepackage{microtype}      
\usepackage{lipsum}
\usepackage{graphicx}
\usepackage{wrapfig}
\usepackage{capt-of}
\usepackage{amsmath}
\graphicspath{{media/}}     

\title{Mage: Cracking Elliptic Curve Cryptography with Cross-Axis Transformers
\thanks{\textit{\underline{Citation}}: 
\textbf{Authors. Title. Pages.... DOI:000000/11111.}} 
}

\author{
  Lily Erickson \\
  EmerGen LLC \\
  Minneapolis, MN\\
  \texttt{lilyerickson@emergenlabs.com} \\
}

\begin{document}
\maketitle

\begin{abstract}
\setlength{\parindent}{0pt}
\setlength{\parskip}{0.5em}
\vspace{1em}
"Cryptography is the art of making discrete data appear statistically random." -Unattributed

\rule{\textwidth}{0.4pt}

With the advent of machine learning and quantum computing, the 21st century has gone from a place of relative algorithmic security, to one of speculative unease and possibly, cyber catastrophe.

Modern algorithms like Elliptic Curve Cryptography (ECC) are the bastion of current cryptographic security protocols that form the backbone of consumer protection ranging from Hypertext Transfer Protocol Secure (HTTPS) in the modern internet browser, to cryptographic financial instruments like Bitcoin.

And there's been very little work put into testing the strength of these ciphers. Practically the only study that I could find was on side-channel recognition, a joint paper from the University of Milan, Italy and King's College, London\cite{battistello2025ecc}.

These algorithms are already considered bulletproof by many consumers, but exploits already exist for them, and with computing power and distributed, federated compute on the rise, it's only a matter of time before these current bastions fade away into obscurity, and it's on all of us to stand up when we notice something is amiss, lest we see such passages claim victims in that process.

In this paper, we seek to explore the use of modern language model architecture in cracking the association between a known public key, and its associated private key, by intuitively learning to reverse engineer the public keypair generation process, effectively solving the curve.

Additonally, we attempt to ascertain modern machine learning's ability to memorize public-private secp256r1 keypairs, and to then test their ability to reverse engineer the public keypair generation process.

It is my belief that proof-for would be equally valuable as proof-against in either of these categories.

Finally, we'll conclude with some number crunching on where we see this particular field heading in the future.
\end{abstract}

\vspace{8em}

\pagebreak

\section{Introduction}
The security of modern ECC algorithms lies in the hardness of the Elliptic Curve Discrete Logarithm Problem (ECDLP).

To keep things simple, the 'irreversibility' of ECC algorithms lies largely in the modulo operand. Each byte cannot exceed a value of 255, and so the large products created by the curve multiplication must necessarily wrap back around to 0 once it exceeds the maximum value allowed by a standard byte. It is very easy to calculate this going forwards.

$$r = a\%n = a-n*\left\lfloor\frac{a}{n}\right\rfloor$$

However it's extremely difficult to reverse this process without knowing the precise quotient (i.e., the number of times the value wrapped around modulo N), and we are already multiplying these values by enormous numbers.

The formula is not without fail though. In order to understand the potential of modern machine learning algorithms in this domain, it is first imperative to explore the two existing vulnerabilities in cryptographic algorithms that we intend to tackle before proceeding.

\rule{\textwidth}{0.4pt}

A cryptographic Side Channel is a data leakage vector (eg: timing attack, power consumption during operations, etc) that exposes information about and compromises the integrity of a cryptographic computation.

\begin{quote}
The Cryptographically Secure Pseudo-random Number Generator (CSPRNG) is considered one of the defacto methods of generating on-the-fly, cryptographically secure information. Normal random number generators are susceptible to timing attacks, however CSPRNG attempts to obfuscate the creation of the initial private key, and in doing so hopes to protect against side-channel and environment based attacks.
	
Machine learning models however, are Universal Function Approximators - there is a very real chance that they are capable of modeling the underlying cipher itself, bypassing this safeguard.

If there is a latent pattern within the ECDLP that a machine learning model could recognize, we should be able to detect it with enough data.
\end{quote}

\rule{\textwidth}{0.4pt}

A Rainbow Table is a partially or completely solved array or mapping of all known input (and/or output) permutations for a given equation.

\begin{quote}
The National Institute of Standards and Technology (NIST) offers guidance that current 256-bit ECC keys have ~128 bits of security, meaning it would take on the order of $2^128$ computational steps to break it.
	
A 256-bit ECC key is a 32-character Uint8Array (each character is an integer between 0-255), meaning that a complete, unoptimized Rainbow Table for all known public -> private key mappings for a 256-bit curve would take up $\sim7.41*10^{57}$ ZB (zettabytes) of data. For comparison, it's estimated that by the end of 2025, earth's datasphere will only be $\sim172$ ZB\cite{idcdata}.
	
However, for any problem in which a solved rainbow table with pre-computed mappings exists, the computation required to traverse and therefore break that algorithm should be O(n).
	
Oh, and did I mention that our friendly Universal Function Approximators are also highly efficient compression algorithms? 
\end{quote}

\rule{\textwidth}{0.4pt}

Hopefully it is obvious why research of this nature is necessary. Humanity is pushing new horizons, and while all eyes are focused on quantum computing, few are eyeballing classical learning algorithms as the potential problem children.

We'll use our Cross-Axis Transformer from our most recently published paper, because it's more computationally efficient\cite{erickson2023cat} than its peers.

Let's continue to our experiment.

\clearpage

\noindent
\begin{minipage}[t]{0.5\textwidth}
\setlength{\parindent}{0pt}
\setlength{\parskip}{0.5em}
\section{Methodology}
\label{sec:methodology}
\subsection{Yes, our model can learn}
We begin this section by running an extremely small training pass of 5,000 key pairs. We wish to intentionally demonstrate that our model can overfit, prior to moving onto the next section.

Working with cryptographically random data has some very unique side effects which might, at first, seem like misconfiguration of the training environment.

After epoch 39, our training accuracy\ref{fig:overfit} was already at 99\%, at which point we stopped the experiment.

\subsection{The problem with Adam}
Adam and AdamW are highly successful and popular optimizers in the world of language modeling. They utilize a series of Moments to help the loss function know when it's moving in the right direction.

CSPRNG is designed to produce cryptographically random values for the private key. When we use these as labels, and average the loss over a large batch size, Law of Large Numbers wins out. Let's check out our training graphs for an AdamW optimizer control group!\ref{fig:adamm}

Look at that loss graph! Because we're optimizing *only* for a cryptographically random set of bytes, there is *no* momentum. Immediately when AdamW picks a direction, it gets told that it's wrong! We've literally broken gradient descent.

And the model is definitely still learning. If the weights weren't updating at all, you'd see repetitive accuracy scores.

It's worth noting that $\frac{1}{256}=0.0039$, which is the accuracy score that indicates our model is truly "guessing" at the labels.

\subsection{The solution}
AdamW is still one of the most successful modern optimizers around.
\begin{equation}
	\theta_{t+1} = \theta_t - \eta \frac{\beta_1 m_{t-1} + (1-\beta_1) g_t}{\sqrt{\beta_2 v_{t-1} + (1-\beta_2) g_t^2} + \epsilon} - \eta \lambda \theta_t
\end{equation}

To correct the error, we disabled the momentum parameter by setting $\beta_1$ to 0.

\end{minipage}%
\hfill
\begin{minipage}[t]{0.45\textwidth}
	\raggedleft
	\centering
	\label{tab:training}
	\begin{tabular}{lr}
		\toprule
		\multicolumn{2}{c}{Table 1: Training Parameters} \\
		\midrule
		Parameter & Value \\
		\midrule
		Model & Cross-Axis Transformer \\
		RoPE Type & 3D \\
		Total Parameters & 784M \\ 
		\midrule
		Hidden Size & 2048 \\
		Number of Layers & 16 \\
		Attention Heads & 16  \\
		\midrule
		Dataset Size & 100,000 \\
		Eval Dataset & 38,200 \\
		\bottomrule
		\vspace{1em}
	\end{tabular}
	\label{fig:overfit}
	\includegraphics[width=0.95\linewidth]{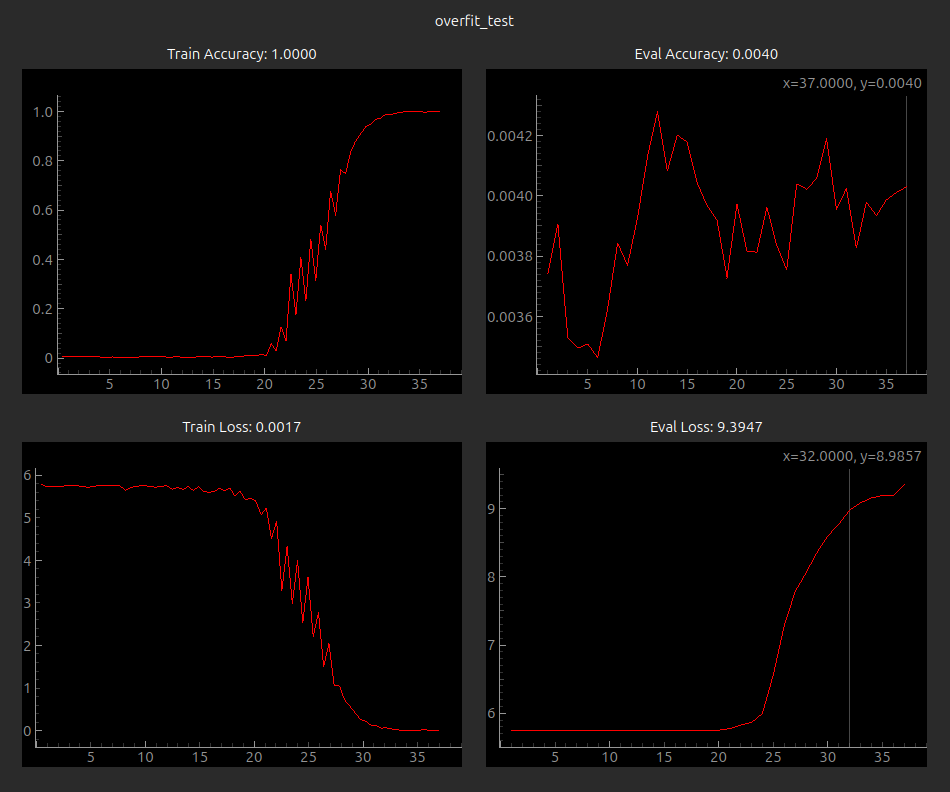}
	\captionof{figure}{An intentional overfit}
	\vspace{1em}
	\label{fig:adamm}
	\includegraphics[width=0.95\linewidth]{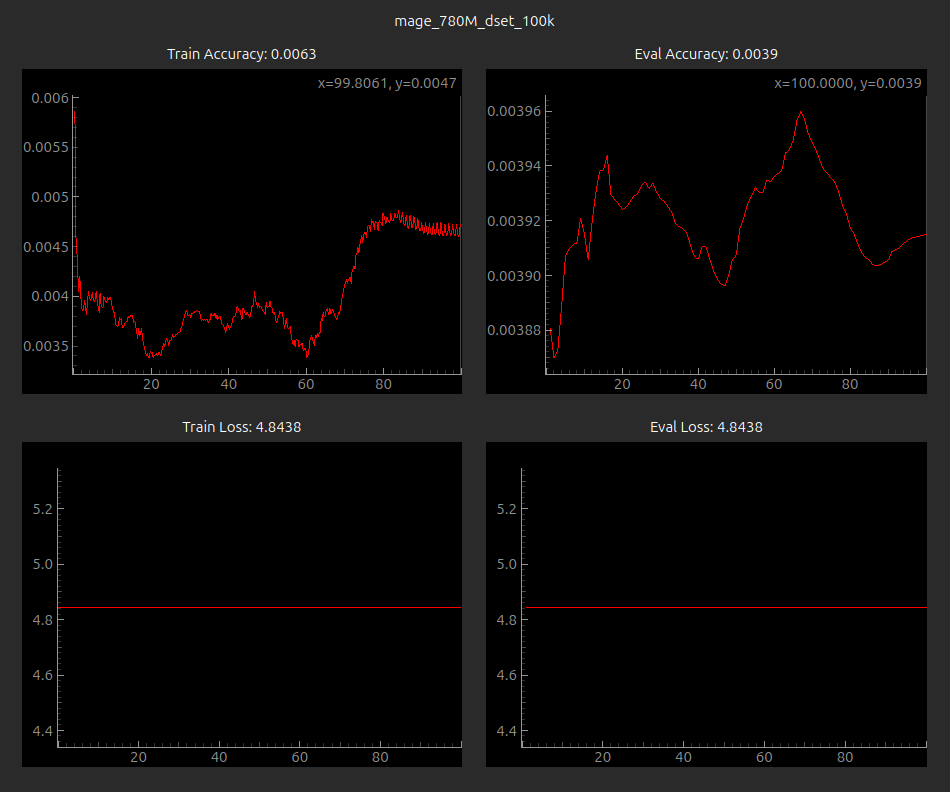}
	\caption{Training with AdamW}
\end{minipage}

\subsection{The goal of our experiment}
We require empirical evidence that our model can memorize its training data at large dataset sizes, and we would also like to see if our model is capable of reverse engineering an ECC private-to-public key generation curve.

\vspace{4em}
\pagebreak

\noindent
\begin{minipage}[t]{0.5\textwidth}
	\setlength{\parindent}{0pt}
	\setlength{\parskip}{0.5em}
	
	\section{Experiments}
	\label{sec:experiments}
	
	We'll start by looking over the results of our first test, using the training dataset described in Methodology\ref{sec:methodology}.
	
	Our model hit 99\% accuracy on the training data by the end of epoch 14, and the eval score had not yet hit some level of statistical significance (~0.0042). Furthermore, as training loss went down, eval loss went up. Since we were not going to see any further improvement, we killed the experiment.
	
	After our first test, it had become clear that our model would make a far better Rainbow Table than an Elliptic Curve simulation. And we were probably going to need a lot more data to tackle the overfitting.
	
	It is at this point that I believe it prudent to mention that the first and second epochs have identical training loss. It is not until the third epoch that the model begins learning a representation of training data and, even the 10M sample sized dataset that we generated but scrapped would be a drop in the bucket of all potential public/private key pairs.
	
	In order to test whether our model is *actually* able to learn a representation of an elliptic curve, we need to replace our dataset with a generator function. Thanks to the sheer size of potential pair candidates, collisions are still almost impossible for our intended purpose. If training loss never increases under this new setup, then we can be thankful that the ECDLP is sound (for now).
	
	Another tweak from our methodology\ref{sec:methodology} baseline is that we are going to disable the Learning Rate Scheduler; a decay function is actually going to hold us back without repeat data.
	
	Training and validation accuracy should be functionally identical now, but the eval accuracy is still averaged over the entire validation step, and is therefore a better statistic.

\end{minipage}%
\hfill
\begin{minipage}[t]{0.45\textwidth}
	\raggedleft
	\centering
	\label{fig:run1}
	\includegraphics[width=0.95\linewidth]{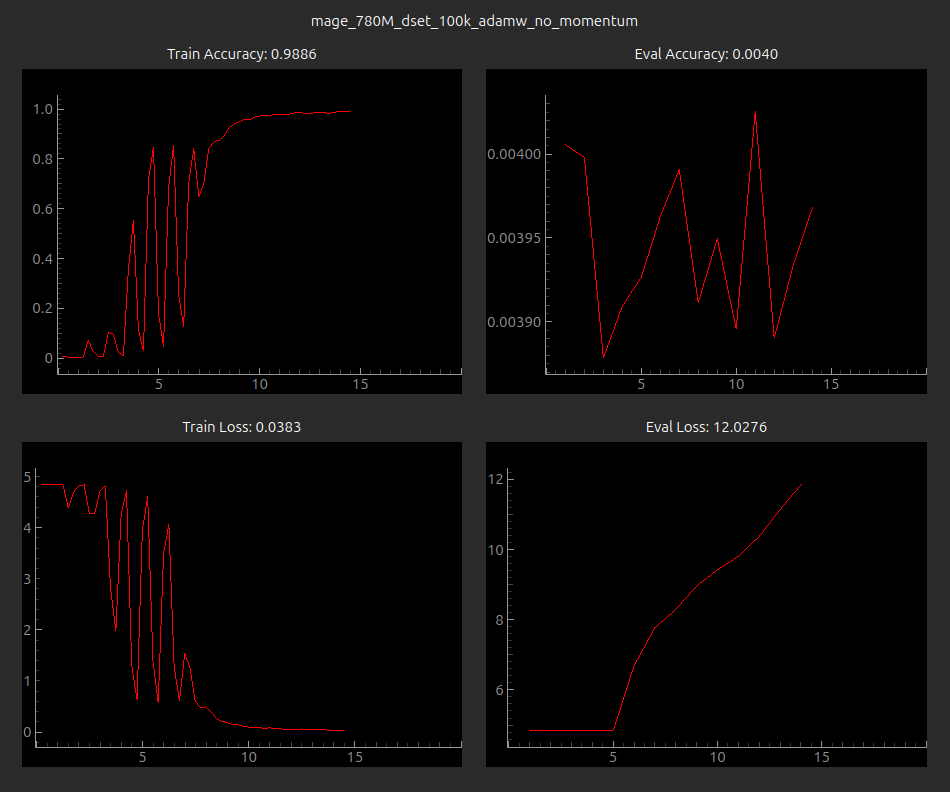}
	\captionof{figure}{The results of our first dataset}
	\vspace{2em}
	\label{fig:run2}
	\includegraphics[width=0.95\linewidth]{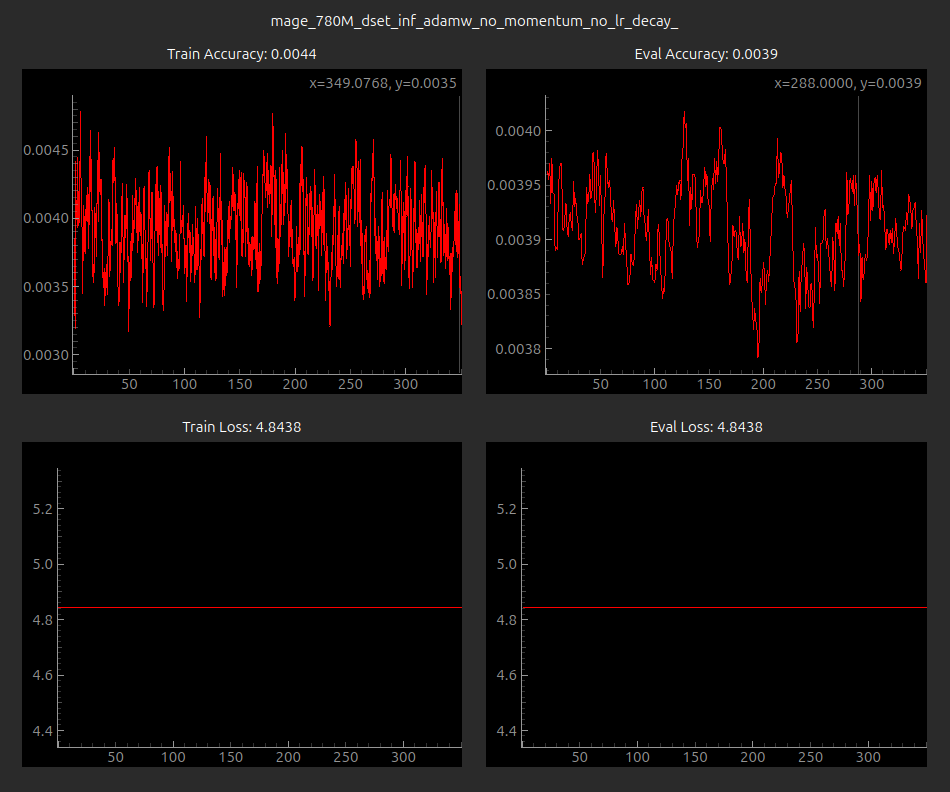}
	\caption{The results of our unbounded dataset}
\end{minipage}

\vspace{1em}

We ran 350 epochs using only randomly generated keypair data with 100,000 samples each, and saw zero improvement in our model. Loss stayed flatline, exactly as we saw in the first two epochs of our smaller training run.

If our CAT is able to learn a mathematical representation of an elliptic curve, it's not going to get there with only 35 million samples. Modern transformers are trained on tens of trillions of tokens, and so our puny hardware is certainly not competing with the big dogs out in the field.

I intentionally refuse to speculate on additional ways to improve or streamline our model architecture or data pipelines for this experiment.

\section{Scaling and Unit Discrepency}
Machine learning model calculations are typically discussed in Floating Point Operations (FLOPs), however cryptographic operations (usually scalar field multiplications and additions) use their own, much heavier operations. We need to bridge these two worlds to have a meaningful discussion on the matter, which actually turns out to be non-trivial.

We acknowledge that these conversions are not ideal. The operations take place on different types of hardware, and we do not have access to a super-computer for benchmarking. Still, we hope that by the end of this process, the implications prove enlightening.

First, let's handle the cryptographic side using some empirical data from the University of Waterloo\cite{cacr2006}, converted to CPU clock cycles.

For the actual number of operations (operation counts) required to create a keypair, we'll use Bernstein and Lange's Explicit-Formulas Database\cite{bernsteinefd}. The creation of a random private key is trivial - we only need to invoke the curve algorithm for the public key.

For the AI conversions, we use simple from-basics derived arithmetic, based on Dual x86 AVX-512 processors that are designed to efficiently handle FLOPs. We assume a 2.5x size backprop gradient for a bf16 precision's combined forward + backwards training pass.

And for completeness' sake, we will include numbers for both our basic baseline CAT, as well as for an open-source, massive language model: Llama 3.1 405B.

\begin{table}[h]
	\centering
	\label{tab:conversions}
	\begin{tabular}{lr}
		\toprule
		\multicolumn{2}{c}{Table 2: Keypair Generation Complexity} \\
		\midrule
		Parameter & Value \\
		\midrule
		Field Multiplications per Keypair & 3,456 \\
		Field Squarings per Keypair & 1,408 \\
		\midrule
		Cycles per Field Mult & 167 (iterations) * 7 (operations) = 1169 cycles \\
		Cycles per Field Sqr & 167 (iterations) * 3 (operations) = 501 cycles \\
		\midrule
		Mult Cycles per Keypair & $4.04*10^6$ \\
		Add Cycles per Keypair & $7.05*10^5$ \\
		Total Cycles per Keypair & $4.75*10^6$ \\
		\bottomrule
	\end{tabular}
\end{table}

\rule{\textwidth}{0.4pt}
\vspace{0em}

\begin{table}[h]
	\centering
	\begin{tabular}{lr}
		\toprule
		\multicolumn{2}{c}{Table 3: Machine Learning Operation Complexity} \\
		\midrule
		Parameter & Value \\
		\midrule
		1x Dual x86 AVX-512 Processor Cycle & 32 DPP FLOPs/core/cycle \\
		Calculations in Bfloat16 Precision & 128 FLOPs/cycle \\
		Epochs to Memorize Keypair & 14 \\
		Bfloat16 training multiplier & 2.5x \\
		Bfloat16 memorization multiplier (single keypair) & 14 * 2.5 / 128 = 0.27 cycles\\
		\midrule
		CAT inference baseline FLOPs & $2.5*10^{10}$ \\
		CAT Total Cycles per Keypair & $6.84*10^9$ \\
		\midrule
		LLaMA‑3 405B inference FLOPs & $1.52*10^{25}$ \\
		LLaMA‑3 405B cycles per keypair  & $4.15625*10^{24}$ \\
		\bottomrule
	\end{tabular}
\end{table}

The above tables\ref{tab:conversions} leave us with two important values: The number of CPU cycles required to generate a secp256r1 keypair, and the number of CPU cycles that were required for our rudimentary model to memorize a keypair. As we push forward, let us keep in mind that although these numbers are useful baselines, they are likely extremely conservative estimates when up against what a thoroughly motivated hacker group or nation-state actor would be able to optimize for.

Of extreme interest, is the fact that it is \textbf{ludicrously efficient} for our model to memorize a single keypair. You can run a batch size of 4 on each core without ever needing to touch quantization.

\section{Cracking the 256-bit secp256r1 curve}
The private key is effectively 32 randomly generated integers between 0 and 255. It is possible to generate every single one deterministically, without any repeats. Compared to what is required to derive the public key, this overhead for private key generation actually is straight up trivial, and we will be ignoring it.

Since there's not consistent guidance on what a cryptographic operation is when estimating resistance, we'll assume the less efficient of the two field operations, and go purely with scalar field multiplication.

For cryptographic resistance, we'll use NIST's common guideline of $2^{128}$ for one keypair. This is the number to beat.

\begin{table}[h]
	\centering
	\label{tab:cycles}
	\begin{tabular}{lr}
		\toprule
		\multicolumn{2}{c}{Table 4: Cycles to Generate an Algorithmic Rainbow Table} \\
		\midrule
		Parameter & Value \\
		\midrule
		NIST's cryptographic resistance (Operations) & $2^{128} = 3.4*10^{38}$ \\
		Cryptographic resistance (cycles) & $3.4*10^{38} * 1169 = 3.4*10^{41}$ \\
		\midrule
		Number of total keypairs & $256^{32} = 1.57*10^{77}$ \\
		Total cycles to generate all keypairs & $1.57*10^{77} * 4.75*10^6 = 7.46*10^{83}$ \\
		Total cycles to memorize all keypairs (CAT) & $1.57*10^{77} * 4.75*10^6 = 1.07*10^{87}$ \\
		Total cycles to memorize all keypairs (Llama3.1 405B) & $1.57*10^{77} * 4.16*10^{24} = 6.53*10^{101}$ \\
		\bottomrule
	\end{tabular}	
\end{table}

And these numbers might look very reassuring if you're not familiar with cryptography or mathematics in general. You might think to yourself that "Yes, Moore's Law gets there eventually but until then we can rest easy."

And to those people, I am sincerely sorry for what I am about to do.

\subsection{The Birthday Paradox}
Any time we are brute-forcing a discrete (albeit quite large) solution state, we need to introduce our favorite cryptographic boogeyman, the Birthday Paradox.

To summarize, as the number of samples in a population increase linearly, their odds of attribute collision increases exponentially. We often use the example of people in a room sharing the same birthday. We'll pull our data from the Handbook of Applied Cryptography's\cite{hac96} Birthday Paradox section.

$$ \text{Odds that at least two people share a birthday: } p = 1 - e^{-\frac{n(n-1)}{730}} $$

\begin{table}[h]
	\label{tab:birthday}
	\centering
	\begin{tabular}{lr}
		\toprule
		\multicolumn{2}{c}{Table 5: The Birthday Paradox} \\
		\midrule
		Number of People & Odds Two Share a Birthday \\
		\midrule
		2 & 0.27\% \\
		5 & 2.7\% \\
		10 & 11.6\% \\
		25 & 56.04\% \\
		\bottomrule
	\end{tabular}	
\end{table}

With only 25 people in the room, the odds that they share a birthday is already well over 50\%. This is because we are not checking if people share a birthday with \textit{you}, but rather if \textit{anybody} shares a birthday with \textit{anybody else}. In cryptography, we often care a lot about avoiding nonce or public key collisions, as these can often immediately invalidate one or more client's security postures.

If we're solving for a 50\% chance of collision, the actual integrity of our security is reduced by the following formula:

$$50\%_p = 1.17 * \sqrt{n}$$

So, let's now give up on trying to memorize every single key, let's instead try to optimize for having a 50\% chance that the client's key resides in our model's memory - the victim can simply flip a coin to decide whether or not they get hacked. We'll recap the key values from the earlier tables to save you some time.

\begin{table}[h]
	\label{tab:cracking}
	\centering
	\begin{tabular}{lr}
		\toprule
		\multicolumn{2}{c}{Table 6: Cracking 256-bit} \\
		\midrule
		Parameter & Value \\
		\midrule
		Base CAT Curve Memorization\ref{tab:cycles} & $1.07*10^{87}$ \\
		Base Llama3.1 405B Curve Memorization\ref{tab:cycles} & $6.53*10^{101}$ \\
		Cryptographic resistance\ref{tab:cycles} & $3.4*10^{38} * 1169 = 3.4*10^{41}$ \\
		\midrule
		$50\%_p$ CAT & $3.83*10^{43}$ \\
		$50\%_p$ Llama3.1 405B & $9.45*10^{50}$ \\
		\bottomrule
	\end{tabular}	
\end{table}

And let's not forget: Llama 3.1 doesn't use Cross-Axis Attention! From our original paper, we're likely looking at an additional computational savings of $1 - \frac{24.85bn}{37.23bn} = 33.25\%$  for free, left completely on the table.

In order to seriously discuss this problem, we are forced to dissect the above equation a bit. We propose two basic formulas for discussing the problem. 
$$
\begin{gathered}
	n = \textbf{target population} \\
	m = \textbf{memorized keys} \\
	a = \textbf{available keyspace ($256^{32}$)} \\
	p = \textbf{solve percentage} \\
	\\
	\textbf{Solving for at least one victim in a given population size: } n = \frac{ln(p)}{ln(1-m/a)} \\
	\textbf{Solving for at least one victim given a percentage accuracy : } p = 1-(1 - \frac{m}{a})^n
\end{gathered}
$$

\section{Conclusion}
Fortunately, based on this research alone as an indicator of where things currently stand, it seems unlikely that a modern machine learning algorithm will not be able to intuitively decipher the Elliptic Curve Discrete Logarithm Problem. Feeding a model cryptographically random information is an amazingly effective way to interfere with these kinds of attempts at to backdoor modern curves and ciphers. Still, it's likely only a matter of time before algorithms like these catch up to us.

On to the scarier half of our experiment, it is clear that our stochastic parrot friends make remarkable rainbow tables. It took only 14 epochs for our model to learn a representation of our training keys. Our model is only $\sim$1.5gb in size, and if not for the fact that training it would take over 50 days on our current hardware, I am confident that our large 10M sample dataset ($\sim$7.5gb in size) would have proven no challenge at all.

If our small 784M parameter CAT model was capable of memorizing enough secp256r1 keypairs to break 50\% of all private keys, it would manage to do so using \textbf{only} the work required to break \textbf{100 keypairs} the traditional way. And remember, this invalidates the \textbf{entire curve}.

To add fuel to the fire, even when compressing the public key, storing a raw mapping of 1 trillion keypairs ($(33+32)*8*10^{12}=2.89^{14}$) would have taken just under 300 terabytes of data. Previously, this meant that our particular attack vector here was a read/write (I/O) problem. It would have been impossible to sample such a table in real time. Here however, you simply need a \textbf{single forward pass} through the AI and you've instantly got your solution. This is a novel brute-force channel over the entire algorithm.

For more practical math, let's assume that an attacker wants to compromise at least one of a generic S\&P500 company's credentials while having access only to their public keys in transit. To do this, they train a model with similar scale and scope to modern open-source AI on the marketplace right now, inline with our Llama 3 example.

Our model can consume significantly more tokens than traditional transformers because it uses a smaller embedding space, more efficient algorithm, and avoids the massive sequence length of language modeling. It also converges much faster than a language model, requiring fewer training epochs to memorize its data.

Additionally, there is no added burden of cleaning and annotating the data. Do not constrain yourself to the "one hundred trillion tokens" count that you might hear thrown around in the LLM world. The following example is akin to what an existing large research laboratory could create within a few months time.

\vspace{1em}

\begin{table}[h]
	\label{tab:breakpoints}
	\centering
	\begin{tabular}{lr}
		\toprule
		\multicolumn{2}{c}{Table 7: Practical Breakpoints} \\
		\midrule
		Parameter & Value \\
		\midrule
		Large Company's Yearly Authentication Requests (Population) & $200,000 * 365 * 500 = 7.3*10^7$ \\
		Has Memorized 1 Quantillion Keypairs (LLM-size) & $10^{18}$ \\
		\midrule
		Current odds of solving a victim's private key & 0.00024\% \\
		\bottomrule
	\end{tabular}	
\end{table}

\vspace{1em}

If we want to assume a sufficiently motivated threat actor or nation-state surveillance organization, pushing the keypair count up to 1 sextillion ($10^{21}$) causes our odds of leaking an employee's access credentials to skyrocket up at \textbf{0.14\%}.

I would also like to briefly touch on the fact that these are permanent solves of existing asymmetric encryption algorithms. Every year that an algorithm is commonly in use, there will be ever-evolving AI dedicated to solving them. Any sufficiently solved asymmetric encryption algorithm is then morphed into a symmetric key. As such, developers should begin to think about systems in which exposing a public key is a matter of last resort.

In the short term, using novel curves or having orders of magnitude more key complexity is a sufficient deterrent to prevent problems.

Long-term, we need solutions in which there is not a perfect mapping between a public and private key.

We are at the frontier of modern AI, and it has immense potential for both good and evil. It is my sincerest hope that everybody is doing everything in their power to keep themselves and their families safe. If you don't use a modern VPN, now is a great time to start.

\section{Ethics}
The results of this experiment were nothing short of disturbing. There are ethical questions to reason with in regards to even publishing a paper on this topic, and the implication that machine learning might be a competitive tool to classical vectors for breaking a single encryption key aside, it looks like the most viable paths forward at this time seem to be in the category of breaking the entire cipher. Quite literally, it is only a matter of time.

This, coupled with the fact that open-source software is typically only important for reproducibility insofar as the required results are - well, difficult to reproduce; we find it prudent to state that we will not be open-sourcing our model architecture, learned weights, or training code for this project at this time. Please see the open-source repository for the original Cross-Axis Transformer\cite{erickson2023cat} if you would like a peak at the original novel model architecture, or our Github repository for the experimental results of this project in CSV format.

\vspace{1em}

\section*{Acknowledgments}
EmerGen is my own personal research organization, which is directly responsible for support, funding, and hardware that powers our current (albeit tiny) experiments.

The last two years have been incredibly stressful for me, as I have been building both a company and restarting a life together at the same time, largely without any help from my peers technologically. I appreciate having the time and inspiration required to put this experiment together while continuing to upgrade our research tech stack.

I am going to end this paper with an open call to any able bodied research institutions to please double check my worth, and to begin running large scale tests.

If you like the direction we are headed and wish to sponsor our work, please reach out to me at lilyerickson@emergenlabs.com

\bibliographystyle{unsrt}  
\bibliography{references}
\end{document}